# Effects of cold atmospheric plasma generated in deionized water in cell cancer therapy

Zhitong Chen, Li Lin, Xiaoqian Cheng, Eda Gjika, Michael Keidar[*]

Department of Mechanical and Aerospace Engineering, The George Washington University,

Washington, DC 20052, USA

**Abstract**

Cold atmospheric plasma (CAP) was shown to affect cells not only directly, but also indirectly by means of plasma pre-treated solution. This study investigated a new application of CAP generated in deionized (DI) water for the cancer therapy. In our experiments, the CAP solution was generated in DI water using helium as carrier gas. We report on the effects of this plasma solution in breast (MDA-MD-231) and gastric (NCI-N87) cancer cells. The results revealed that apoptosis efficiency was dependent on the plasma exposure time and on the levels of reactive oxygen and nitrogen species (ROS and RNS). The plasma solution that resulted from 30-minute treatment of DI water had the most significant effect in the rate of apoptosis.

---

[*] Corresponding Author:
E–mail address: zhitongchen@gwu.edu, keidar@gwu.edu

## 1. Introduction

Plasma is a fully or partially ionized gas consisting of positive and negative ions, free electrons, free radicals, ozone and ultraviolet radiation[1,2]. Historically, plasma could be generated only at high temperatures or in vacuum[3,4], while more recent studies have reported on plasma generated at atmospheric pressure and at room temperature. This type of plasma is referred to as non-thermal plasma or cold atmospheric plasma (CAP)[5]. The potential applications of CAP have created a new field of study known as plasma medicine[6]. One of the main benefits of the CAP technology includes its ability to offer a minimally-invasive surgery option that allows for selective cell death without influencing the healthy tissue. Some of the applications of CAP include: sterilization of infected tissue, inactivation of microorganisms, wound healing, skin regeneration, blood coagulation, tooth bleaching, and cancer therapy[6-11]. According to several studies, the main therapeutic effects of this technology are linked to presence of reactive oxygen species (ROS) and reactive nitrogen species (RNS) generated from the plasma[10,12]. The dose of the plasma species delivered to target tissue site are determined by plasma experimental parameters. Wiseman et. al. has reported that extreme amounts of reactive species (RS) are able to induce cell apoptosis while damaging proteins, lipids, and DNA[13]. Additional studies have reported on cancer cell effects from direct and indirect treatment through plasma irradiated medium[14-16]. While effects of plasma-stimulated medium in cancer cells and microbial inactivation are under current investigation[17,18], the effects of helium plasma generated in DI water are yet to be evaluated.

In this study, we have designed a CAP device submerged in DI water using helium as feeding gas. The plasma solution produced for this device is also referred to as plasma generated in DI water. This plasma solution has the potential to be utilized as an oral medicine or to even be paired with other drugs or used as standalone drug and injected into tumors. Here, we have investigated the effects of the plasma generated in DI water for 5, 10, 20, and 30 min on human breast and gastric cancer cells. The voltage/current, optical emission spectrum and discharge density of helium plasma were characterized. The ROS/RNS concentrations of plasma solutions and relative metabolic activity of breast and gastric cancer cells were also determined.

## 2. Materials and Methods

The human breast cancer cell line, MDA-MB-231, was provided by Dr. Zhang's lab at the George Washington University. MDA-MD-231 cells were cultured in Dulbecco's Modified Eagle Medium (DMEM, Life Technologies) supplemented with 10% (v/v) fetal bovine serum (Atlantic Biologicals) and 1% (v/v) penicillin and streptomycin (Life Technologies). The human gastric cancer cell line, NCI-N87, was purchased from American Type Culture Collection (ATCC). NCI-N87 cells were cultured in RPMI-1640 Medium (ATCC® 30-2001™) supplemented with 10% (v/v) fetal bovine serum (Atlantic Biologicals). Both cell cultures were maintained at 37 °C in a humidified incubator containing 5% (v/v) $CO_2$. All fluorescence and absorbance measurements were recorded with the Synergy H1 Multi-Mode plate reader. The results for the mean ± standard deviation were plotted with Origin 8. A student t-test was performed to check for statistical significance (*p<0.05, **p<0.01, ***p<0.001). The sample size for all experiments was n=3.

The cells were plated in 96-well flat-bottom plates at 3000 cells/well in 70 $\mu$L of complete cell culture medium. The cells were incubated for 24 hours to ensure proper cell adherence and stability. Cell confluency was confirmed at about 40%. Cells were further incubated at 37 °C for 24 and 48 hours in the presence of 30 $\mu$l of DMEM, RMPI, untreated DI water (0 min), and plasma solutions. The relative metabolic activity of the breast cancer cells was measured for each incubation time point (24 and 48 hours) with an MTT assay. 100 $\mu$L of MTT solution (3-(4, 5-dimethylthiazol-2-yl)-2,5-diphenyltetrazolium bromide) (Sigma-Aldrich) was added to each well followed by a 3-hour incubation. The MTT solution was discarded and 100 µL/well of solvent (0.4% (v/v) HCl in anhydrous isopropanol) was added to each well. The absorbance was recorded at 570 nm with microplate reader. The relative metabolic activity of the gastric cancer cells was measured with a Cell Counting Kit 8 assay from Dojindo Molecular Technologies, MD. After each incubation time point (24 and 48 hours), the original culture medium was aspirated and replaced with 10 µL/well of CCK 8 reagent. The plates were incubated for 3 hours at 37 °C. The absorbance was measured at 450 nm using a microplate reader.

A Fluorimetric Hydrogen Peroxide Assay Kit (Sigma-Aldrich) was used for measuring the levels of $H_2O_2$ in the plasma solution. A detailed protocol can be found on the Sigma-Aldrich website. Briefly, we added 50 µl of standard curve samples, controls, and experimental samples to a black 96-well flat-bottom plates. Additionally, we added 50 µL of Master Mix to each of wells. The plates were incubated at room temperature for 30 minutes protected from light. Fluorescence was measured with a microplate reader at Ex/Em: 540/590 nm.

RNS level in plasma solution were determined with the Griess Reagent System (Promega Corporation) according to the instructions provided by the manufacturer. The absorbance was measured at 540 nm with a microplate reader. Hydroxyl free radical levels were measured by utilizing a methylene blue solution. 0.01g/L of methylene blue (MB) in DI water was treated with plasma for 0, 5, 10, 20, and 30 min. 100 µL/well of these solutions were added in triplicate to a black 96-well clear bottom plate. The absorbance was measured at 664 nm with a microplate reader. UV-visible-NIR in the wavelength range of 200-850 nm, was used to detect and investigate presence of various RNS and ROS (nitrogen [$N_2$], nitric oxide [–NO], nitrogen cation [$N^{+2}$], atomic oxygen [O], and hydroxyl radical [–OH]) in plasma generated in DI water. The spectrometer and the detection probe were purchased from Stellar Net Inc. The optical probe was placed in front of the plasma jet nozzle at a distance of 3.5 cm. The integration time for data collection was set to 100 ms.

## 3. Results and Discussions

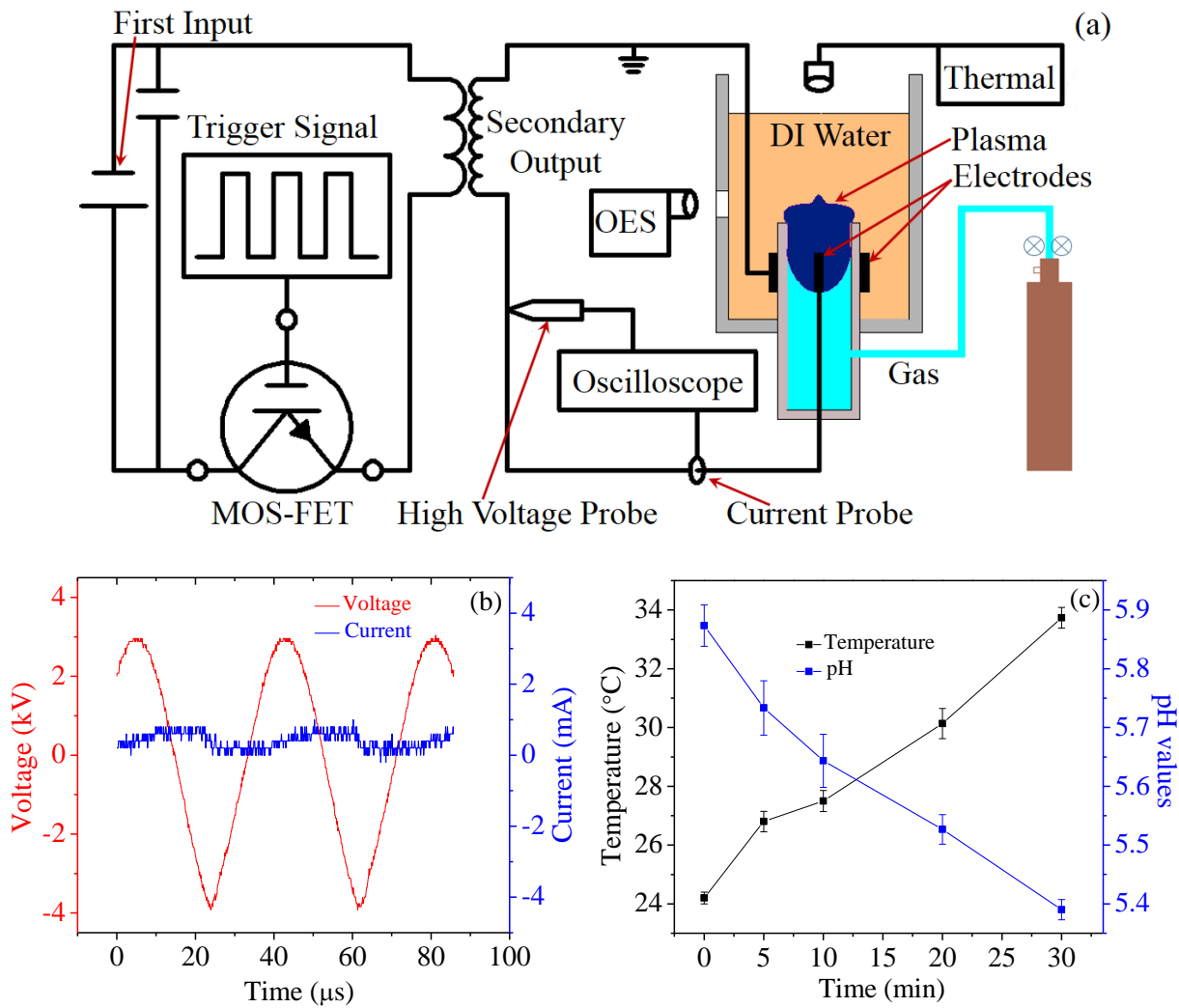

Fig. 1. (a) Diagram of the CAP device setup consisting of a HV pulse generator connected to a pin-to-plate electrode system submerged in deionized water. (b) Measured voltage and current generated by the CAP device. (c) Temperature and pH changes of plasma solutions based on treatment time.

The CAP device submerged in DI water is shown in Fig. 1a. The device consisted of 2 electrodes assembled with a central powered electrode (1 mm in diameter) and a grounded outer electrode wrapped around the outside of the quartz tube (4.5 mm in diameter). The electrodes were connected to a secondary output high voltage transformer. The graphs of CAP for current and voltage are shown in Fig. 1b. The peak-peak voltage was about 7 kV and the average current was about 0.40 mA. The frequency of the discharge generated in DI water was around 25 kHz. Industrial grade helium with a flow rate of 0.4 L/min was used for testing. The plasma produced inside DI water generated four plasma solutions after 5, 10, 20 and 30 minutes of treatment. Fig. 1c shows that the temperature of plasma solution increases while pH decreases with treatment time.

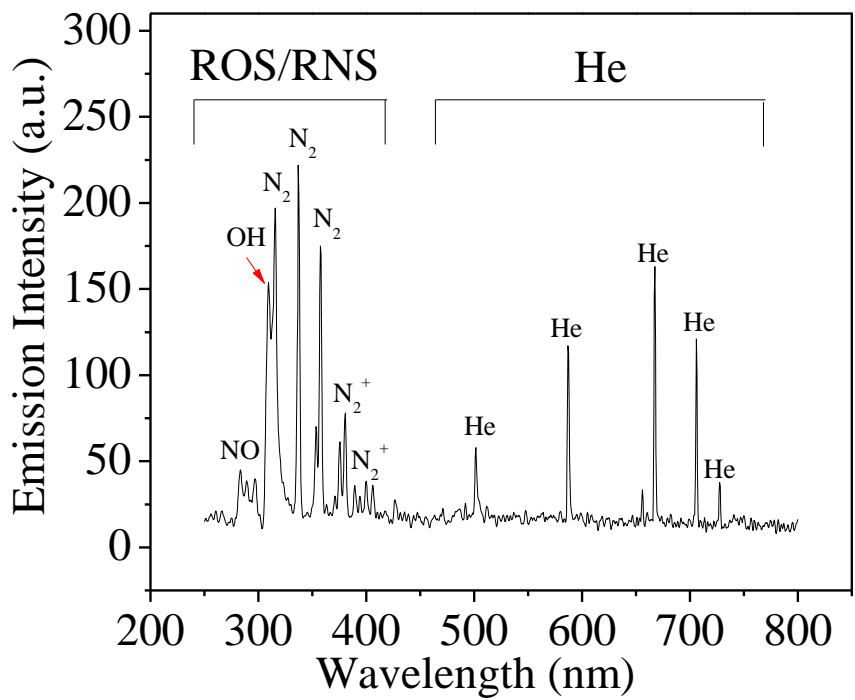

Fig. 2. Optical emission spectrum detected from the plasma device submerged in DI water in the 250-800 nm wavelength range.

The reactive species produced from the CAP device in DI water are shown in Fig. 2. The identification of the emission bands was performed according to the reference[19]. In the 250-300 nm wavelength range, weak emission bands were detected as NO lines[20]. The emission bands between 300 and 500 nm have still not been clearly identified in the literature[21]. However, we anticipated that OH was present at 309 nm, while the wavelength of 337, 358, and 381 nm could be indicative of the low-intensity $N_2$ second-positive system ($C3\Pi u - B3\Pi g$). The bands between 250 nm and 425 nm could be defined as ROS/RNS. The helium bands were assigned between 500 and 750 nm as shown in Fig. 2.

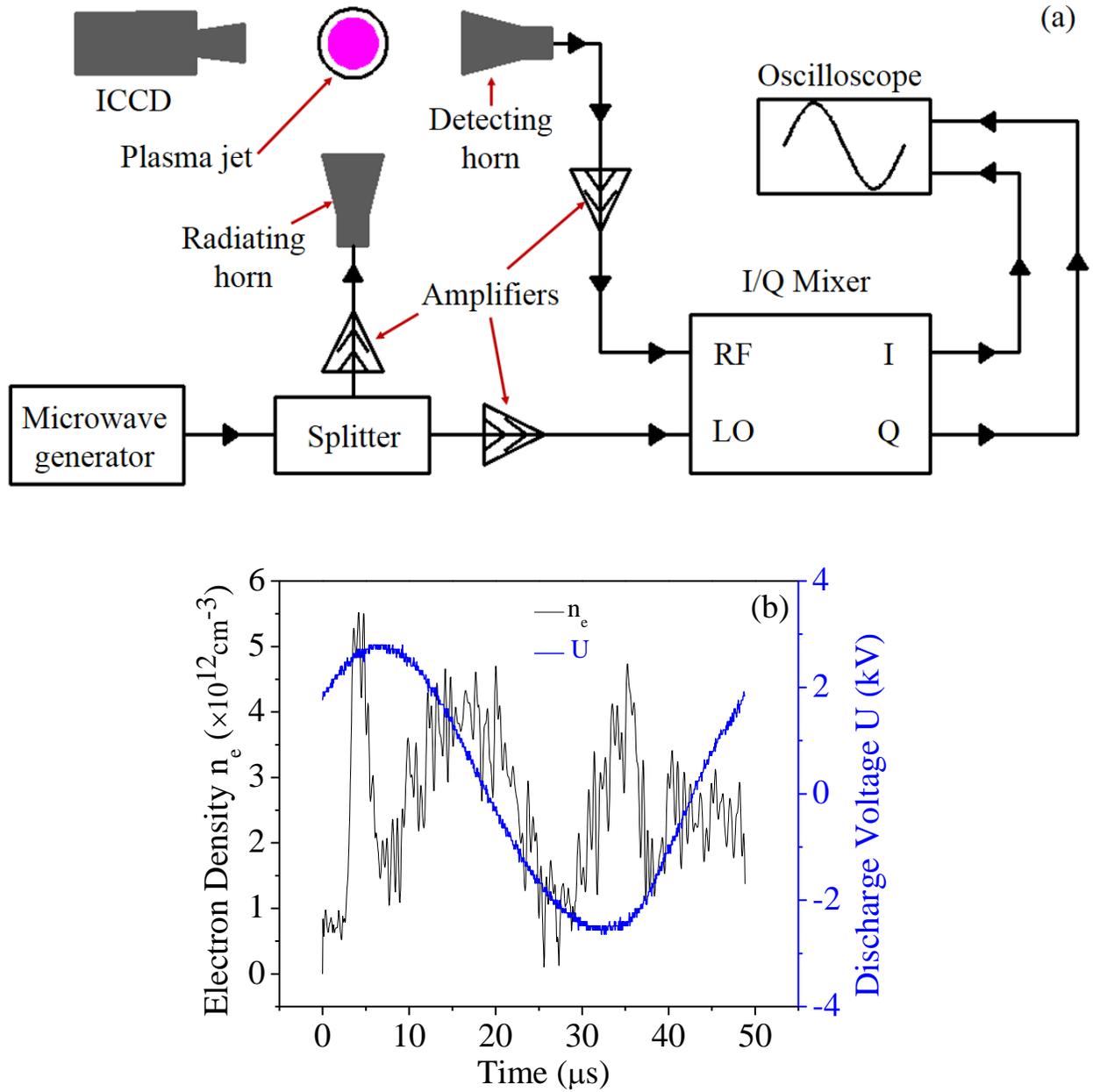

Fig. 3. (a) The schematics of the RMS experimental setup. (b) Temporal evolution of plasma density in helium cold atmospheric plasma jet in air

The Rayleigh microwave scattering system (RMS) was used to determine the intensity of the electrons. RMS consisted of two microwave horns that were used for the radiation and detection of the microwave signal as shown in Fig. 3a. The scattered signal was measured after the linearly polarized microwave radiation was scattered on the collinearly-oriented plasma channel. A homodyne I/O Mixer providing in-phase (I) and quadrature (Q) outputs was used to detect the scattered signal. For the entire range of scattered signals, the amplifiers and mixer were operated in linear mode. Plasma density was obtained from plasma conductivity by using the following expression: $\sigma\ (\Omega^{-1} cm^{-1}) = \frac{2.82 \times 10^{-4} n_e v_m}{(w^2 + v_m^2)}$, where $v_m$ is the frequency of the electron-neutral collisions, $n_e$ is the plasma density, and $w$ is the angular frequency[22]. Plasma conductivity can be expressed as $U = A\sigma V$, where $A = 263.8\ V\Omega/cm^2$, $U$ is the output signal and $\varepsilon$ is the scatter constant[23]. The volume of the plasma column was determined from the intensified charged-coupled device (ICCD) images. The radius ($R$) of the streamer column was determined from the size of the central highly luminous filament. Temporal evolution of plasma density is presented in Fig. 3b with an average electron density of $2.56 \times 10^{12}$ cm$^{-3}$. The waveform of the discharge voltage has been inserted in Fig. 2(b). Fig. 3b shows three electron density peaks per one discharge period that correspond to the discharge voltage. The electron density and its discharge voltage waveform in one complete discharge period are aligned with regard to the x-axis (time). The temporal fluctuation and discharge voltage (Fig. 3b) were different from the power source frequency (Fig. 1c) because the system's hardware worked in air resulting into discharge frequency and voltage alterations. The circuit diagram in Fig. 1a indicates that the pair of electrodes are actually a capacitive load of the transformer. Therefore, a different capacitance between the electrodes could result into different discharge voltage and frequency.

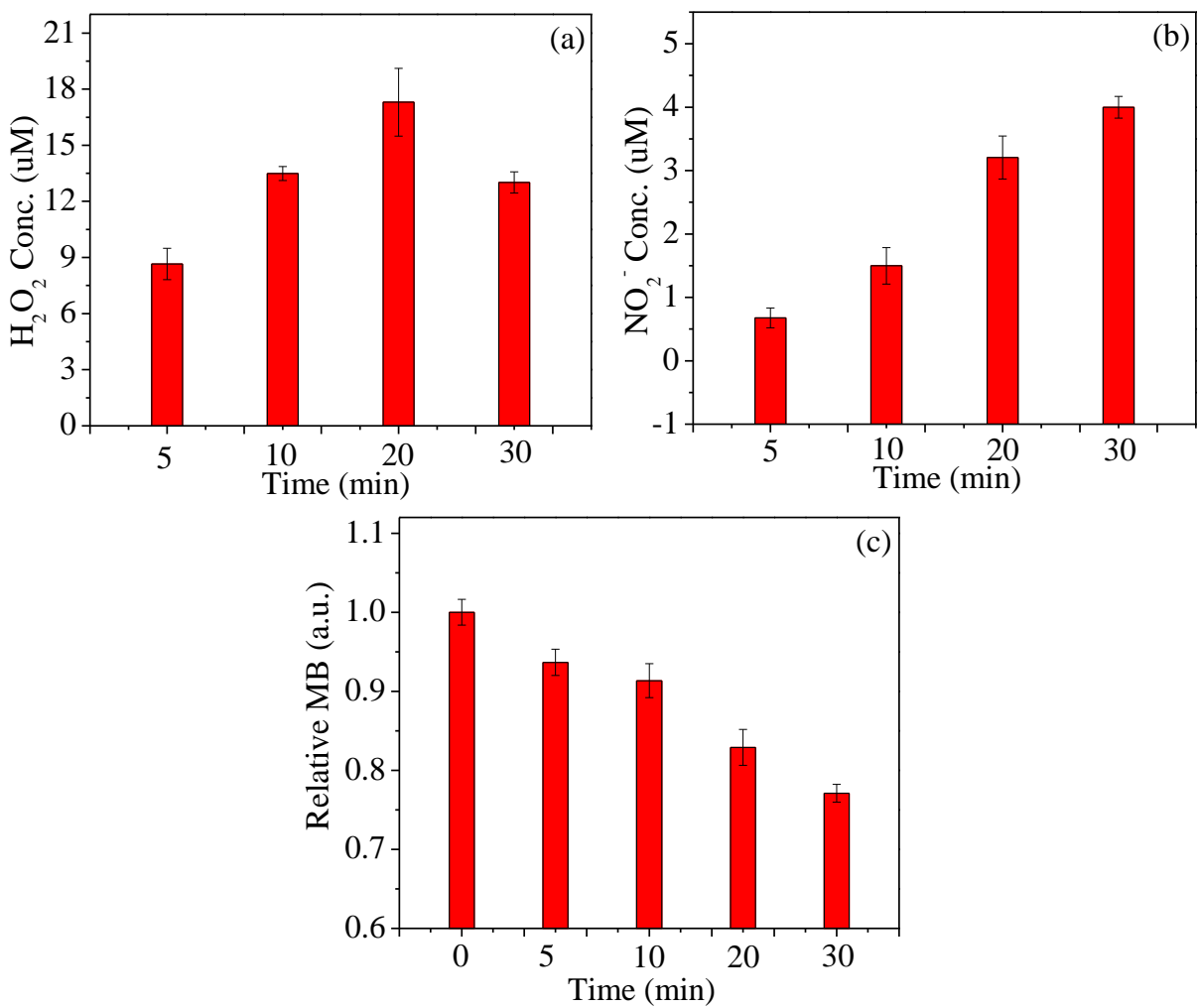

Fig. 4. (a) $H_2O_2$ and (b) $NO_2^-$ concentrations in plasma generated DI water using Helium as carrier gas, and (c) relative methylene blue (hydroxy radicals), (n=3)

CAP can produce chemically reactive species in DI water. Many of the radicals generated during the discharge can contribute to complex reactions[24]. These reactions result in the formation of other short- and long-lived radicals or species. $H_2O_2$ and $NO_2^-$ are relatively long-lived species in plasma solution. The $H_2O_2$ concentration in the plasma solution produced by the DI water submerged CAP device is shown in Fig. 4a. The concentration of $H_2O_2$ increased with treatment time up to 20 minutes. While between 20 and 30 minutes the concentration decreased. The methylene blue (MB) was used to quantify the generation of hydroxyl free radicals (•OH), (Fig. 4c). It is well established that MB reacts with the hydroxyl free radicals (•OH) of the aqueous solution by resulting in a visible color change. $H_2O_2$ is produced in DI water within a few microsecond from hydroxyl (•OH)[25]. A description of the possible mechanism of $H_2O_2$ formation can be found in one of our recent publications[26].

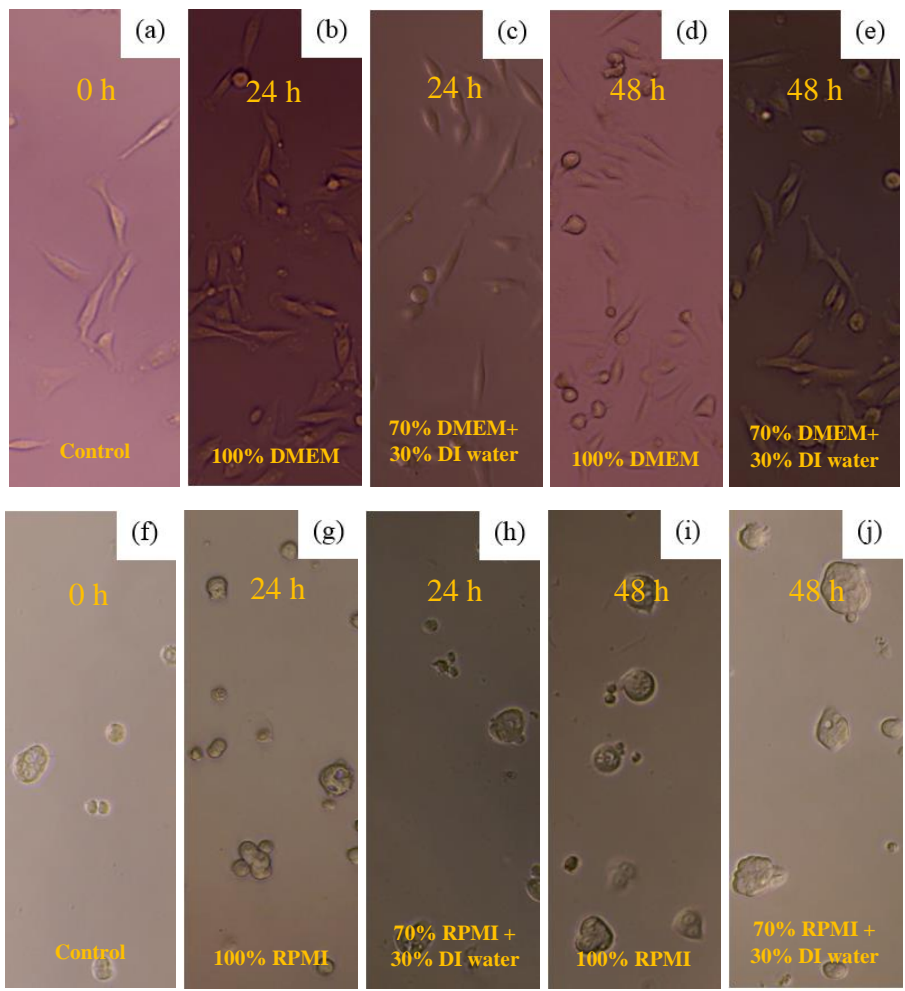

Fig. 5 The morphology of (a, b, c, d, and e) breast cancer cell and (f, g, h, i, and j) gastric cancer cells. (a) and (f) are 0 h with 100% DMEM and RPMI, respectively. (b) 100% DMEM, (c) 70% DMEM + 30% DI water, (g) 100% RPMI, and (h) 70% RPMI + 30% DI water after a 24-hour incubation. (d) 100% DMEM, (e) 70% DMEM + 30% DI water, (i) 100% RPMI, and (j) 70% RPMI + 30% DI water after a 48-hour incubation.

In this study, the plasma solution was obtained from CAP generated in DI water with 5 min, 10 min, 20 min, and 30 min treatment time. Hydrogen peroxide is thermodynamically unstable and decomposes to form water and oxygen. The rate of decomposition increases with rising temperature[27]. Fig. 1c. shows that the temperature of plasma solutions increases with treatment time. These results may explain the decrease in $H_2O_2$ concentration that occurs after 20 min. The concentration of $NO_2^-$ continuously increases with treatment time as shown in Fig. 4b. The $NO_2^-$ mainly originates as NO, while most of NO is formed in the gas phase during the afterglow a few milliseconds after the discharge pulse. It is known that $NO_2^-$ is a primary breakdown product of NO in DI water[28]. Due to DI water contact with air, it is possible that $O_2$ and perhaps $N_2$ are coming from air. Another possibility is that $N_2$ is from the industrial grade helium.

We foresee that in future, the plasma solution might be paired with drugs which assist against cancers of the digestive system. In this study, we not only show the effects of DI water treated by CAP, we have also investigated plasma generated in RPMI and DMEM. However, we have found that plasma generated in RPMI and DMEM is very unstable, which lead to inaccurate results. Furthermore, it is very difficult to directly treat blood with plasma due to the fact that blood coagulates and it has a higher viscosity coefficient than water. Thus, we can use the strategy of plasma generated in water for injecting it into the bloodstream around breast area. Fig. 5 shows the morphology of breast cancer cells and gastric cancer cells with and without 30% DI water at 24 and 48 hour incubations. The cells with and without DI water appear morphologically similar for both incubation time points. Fig. 6 shows the relative metabolic activity of the human breast and gastric cancer when they were exposed to DMEM/RPMI, untreated DI water (0 min) and plasma solutions (5, 10, 20, and 30 mins) for 24 and 48 hour incubations.

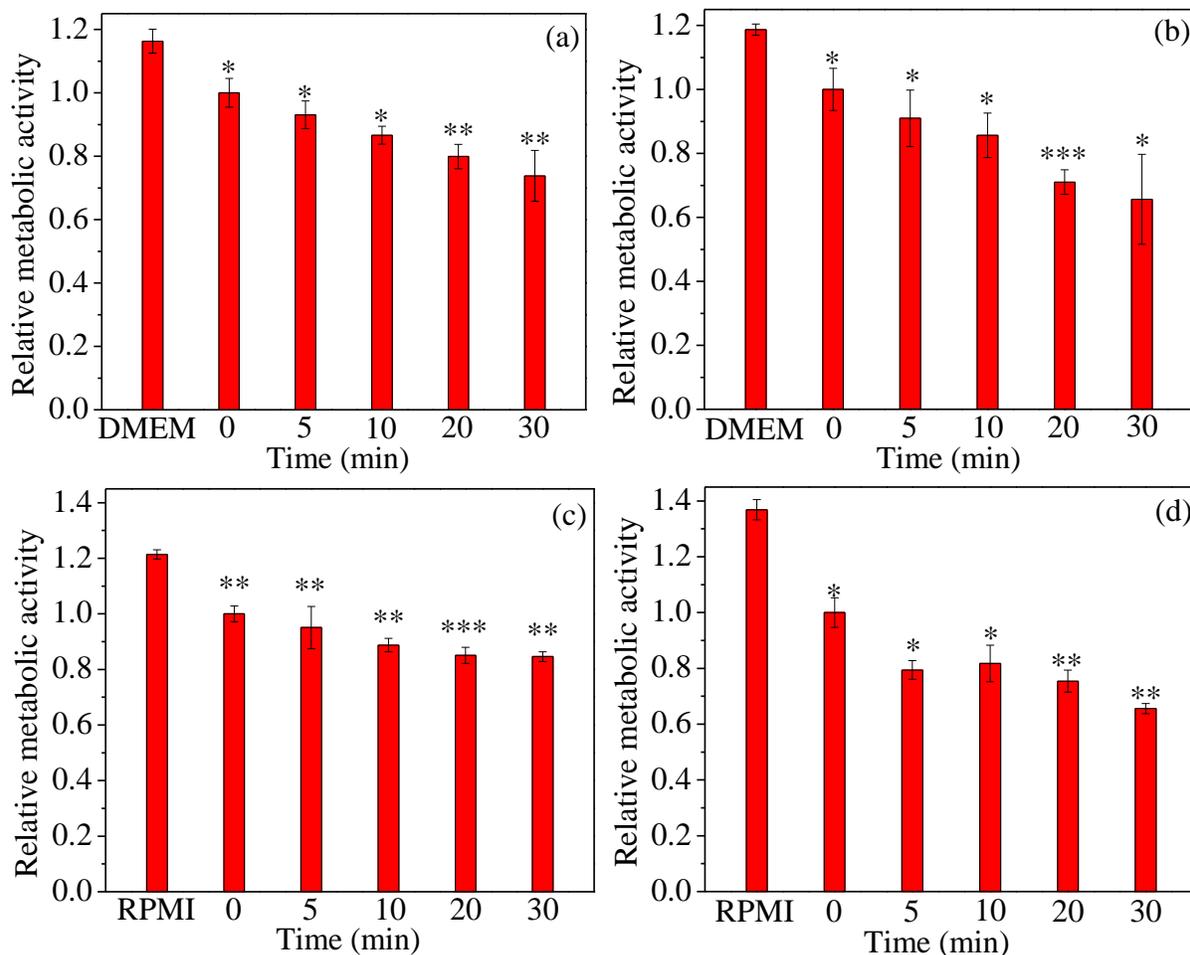

Fig. 6. The effects of the five solutions: DMEM/RPMI, untreated DI water (0 min), and plasma solutions generated in DI water during 5, 10, 20, and 30 min treatment. Cell relative metabolic activity of the human breast cancer cells (MDA-MB-231) at (a) 24-hour incubation and (b) 48-hour incubation. Cell relative metabolic activity of human gastric cancer cells (NCI-N87) at (c) 24-hour incubation and (d) 48-hour incubation. The ratios of surviving cells for each cell line were calculated relative to controls (DI water (0 min)). Student t-test was performed, and the statistical significance compared to cells present in DMEM/RPMI (first bar) is indicated as *$p < 0.05$, **$p < 0.01$, ***$p < 0.005$. (n=3)

Initially, we measured the cell relative metabolic activity using the same assay for both cell lines. However, during our testing, we found that MTT was better for quantifying changes in the breast cancer cells, while Cell Counting Kit 8 assay was more effective for gastric cancer cells. DMEM, RPMI, and untreated DI water (0 min) were used as controls to normalize all data for the cell relative metabolic activity. Fig. 6a and Fig. 6c show that after 24 hours, the relative metabolic activity of the breast and gastric cancer cells treated with DI water (0 min) decreased to 14.0% and 17.6% in comparison to the DMEM/RPMI control condition. The relative metabolic activity of the breast and gastric cancer cells treated by plasma solution was lower than that of the untreated DI water (0 min) and dropped with increasing treatment time. After 48 hours of incubation, the cell relative metabolic activity of the breast cancer cells (compared with DMEM) decreased by approximately 23.4%, 27.8%, 40.1%, and 44.7%, respectively, according to 5, 10, 20, and 30 min treatment duration (Fig. 5b). Whereas the relative metabolic activity of the gastric cancer cells (compared with RPMI) decreased by 42.0%, 40.3%, 45.0%, and 52.1%, respectively, according to treatment durations ranging from 5-30 min. (Fig. 5d). The most significant effect based on the relative metabolic activity was observed for the 30 min plasma treated solution.

A decrease in cell relative metabolic activity was accompanied with an increase in the concentration of $NO_2^-$ and $H_2O_2$. ROS and RNS are known to induce cell proliferation as well as cell death. ROS are known to induce both apoptosis and necrosis[29], while RNS can induce cell death via damage DNA[30]. Our results in Fig. 3 show that the ROS concentration is highest at 20-minute treatment while the RNS concentration is highest at 30-minute. The trend of cell death can be attributed to the increase of RNS concentration with treatment time. A synergistic effect of RNS and ROS is suspected to play a key role in the apoptosis of the plasma solutions. $NO_2^-$ concentration increased with treatment time, while $H_2O_2$ concentration decreased after 20 minutes of plasma treatment. Thus, no $H_2O_2$ might be present in solutions treated for longer than 30 min, therefore those types of plasma solutions might not be useful.

4. **Conclusions**

In summary, cold atmospheric plasma was generated in DI water using helium as carrier gas after 5, 10, 20, and 30 min. ROS concentration initially increased with treatment time and then decreased after 20 min, while RNS concentration continually increased with treatment time. A synergistic effect of RNS and ROS present in the plasma solution is suspected to play a key role in the rate apoptosis. The plasma generated in DI water during a 30-minute treatment had the most significant affect in inducing apoptosis in both breast and gastric cancer cells.

**Acknowledgement:** This work was supported in part by The National Science Foundation, grant #1465061. We thank Dr. Ka Bian and Dr. Ferid Murad from the Department of Biochemistry and Molecular Medicine at The George Washington University for their support with the experiments for measuring ROS and RNS. We also thank Dr. Grace Zhang from the Department of Mechanical and Aerospace Engineering at The George Washington University for providing the breast cancer cell line.